\documentclass[fleqn,10pt]{wlscirep}
\usepackage[utf8]{inputenc}
\usepackage[T1]{fontenc}
\usepackage{bm}
\usepackage{siunitx}
\title{Project 1000 x 1000: Centrifugal melt spinning for distributed manufacturing of N95 filtering facepiece respirators}
\author[1]{Anton Molina$^\dag$}
\author[2]{Pranav Vyas$^\dag$}
\author[3]{Nikita Khlystov$^\dag$}
\author[2]{Shailabh Kumar}
\author[2]{Anesta Kothari}
\author[4]{Dave Deriso}
\author[5]{Zhiru Liu}
\author[2]{Samhita Banavar}
\author[6]{Eliott Flaum}
\author[2]{Manu Prakash*}
\affil[1]{Department of Materials Science and Engineering, Stanford University, Stanford, CA 94305}
\affil[2]{Department of Bioengineering, Stanford University, Stanford, CA 94305}
\affil[3]{Department of Chemical Engineering, Stanford University, Stanford, CA 94305}
\affil[4]{Department of Electrical Engineering, Stanford University, Stanford, CA 94305}
\affil[5]{Department of Applied Physics, Stanford University, Stanford, CA 94305}
\affil[6]{Program in Biophysics, Stanford University, Stanford, CA 94305}
\affil[*]{\textit{To whom correspondence should be addressed; E-mail:} manup@stanford.edu}
\affil[$^\dag$]{\textit{These authors contributed equally}}

\begin{abstract}
The COVID-19 pandemic has caused a global shortage of personal protective equipment. 
While existing supply chains are struggling to meet the surge in demand, the limited supply of N95 filtering facepiece respirators (FFRs) has placed healthcare workers at risk.  This paper presents a method for scalable and distributed manufacturing of FFR filter material based on a combination of centrifugal melt spinning utilizing readily available cotton candy machines as an example. The proposed method produces nonwoven polypropylene fabric material with filtering efficiency of up to 96\% for particles 0.30-0.49~\si{\micro\meter} in diameter. 
We additionally demonstrate a scalable means to test for filtration efficiency and pressure drop to ensure a standardized degree of quality in the output material. We perform preliminary optimization of relevant parameters for scale-up and propose that this is a viable method to rapidly produce up to one million N95 FFRs per day in distributed manner with just six machines per site operating across 200 locations. We share this work as a starting point for others to rapidly construct, replicate and develop their own affordable modular processes aimed at producing high quality filtration material to address the current FFR shortage globally.

\end{abstract}

\begin{document}

\flushbottom
\maketitle

\thispagestyle{empty}

\section*{Introduction}
The COVID-19 pandemic caused by the SARS-CoV-2 coronavirus has resulted in widespread shortages of personal protective equipment, especially N95 filtering facepiece respirators (FFRs), which are critical to the safety of patients, caretakers, and healthcare workers exposed to high volumes of aerosolized viral pathogens. The unprecedented demand for N95 FFRs has rapidly depleted existing supply chains, causing medical centers in major cities to initiate efforts to decontaminate N95 FFRs for reuse. Resource limited regions that normally have very limited access to N95 FFRs have limited choice but to utilize ineffective substitutes such as T-shirts and tissues, which places their already-limited number of healthcare workers at great risk of infection. While existing N95 supply chains have struggled to meet the surge in demand, bad actors have begun flooding the market with counterfeit FFRs that are incapable of providing appropriate respiratory protection \footnote{At the time of writing, the CDC is actively maintaining a growing \href{https://www.cdc.gov/niosh/npptl/usernotices/counterfeitResp.html}{ list of counterfeit N95 products}.}.
For these reasons, there is an urgent need to augment the existing supply chain of valid N95 FFRs through distributed, rapidly accessible manufacturing methods combined with quality control and local testing and validation techniques. 

The National Institute for Occupational Safety and Health (NIOSH) require  filters with an N95 rating to remove at least 95\% of particles $\ge 0.3$ \si{\micro\meter} in size \cite{national1996niosh}. Commercial N95 filters typically consist of three to four layers of non-woven fibrous material that trap aerosolized viral particles within its fiber matrix using a combination of inertial and electrostatic forces.
The fiber matrix is typically composed of a blend of nano- and micrometer diameter polymer-based fibers. Previous studies\cite{kim2007direct, kim2010application, bonilla2012direct} using electrostatic field meters suggest\footnote{There's limited publicly-available data on the electrostatic charges of commercial polymers, and the measurement process itself is quite involved.} that these fibers hold a baseline electrostatic potential of up to \si{10 \volt} at a tip-sample distance of 75~\si{\nano\meter}.

The N95 filter material is produced on an industrial scale in a process called "meltblowing" \cite{huang2017, huang2019}, where high velocity air streams are blown through nozzles that coaxially extrude molten polymer. Thermoplastics such as polypropylene and poly-4-methyl-1-pentene are commonly used in this process because of their low water retention and desirable melt-flow properties \cite{drabek2019}. Electrostatic charge contributes as much as 95\% of the filtration efficiency \cite{tsai2020} and is typically accomplished using corona charging \cite{klaase1984}. However, expanding production to new industrial meltblowing facilities is a major effort that includes precision manufacturing of large-scale, specialized extrusion dies as well as design of an extensive fabrication workflow, requiring months of construction before operation is possible \cite{reifenhauser2020}. Existing N95 FFR manufacturing methods by meltblowing therefore are incompatible with the urgent need for increased supply.

In this study, we investigate centrifugal melt spinning (CMS) as a small-scale, distributed manufacturing approach for N95 FFR production. With very simple equipment and operational methods, we hypothesize that CMS can be rapidly deployed to address the increased demand in N95 FFRs. Laboratory-scale CMS-based methods are estimated to have 50-fold greater throughput than equivalent electrospinning methods, with production rates of up to 60 \si[per-mode=symbol]{\gram\per\hour} per orifice \cite{rogalski2017}. The physical footprint of CMS operation is significantly reduced compared to that of industrial meltblowing operations, allowing multiple CMS production lines to be run in parallel at smaller, local settings. CMS-based methods have been previously used for the fabrication of poly(lactic acid), poly(ethylene oxide), and polypropylene nano- and microfibers \cite{raghavan2013, parker2010}.

Here, we apply centrifugal melt spinning to address the ongoing N95 FFR shortage and construct a readily distributable, low-cost and modular laboratory-scale fiber production apparatus. We report characterization of fiber morphology and filtration performance of CMS-produced polypropylene fiber material and investigate electrostatic charging through the application of an external electric field in two different ways. We perform preliminary optimization of parameters relevant to N95 FFR manufacturing using CMS and propose how the process may be scaled up through distributed manufacturing.

\section*{Results and Discussion}

\begin{figure}[!ht]
\begin{center}
\includegraphics[width=0.75\linewidth]{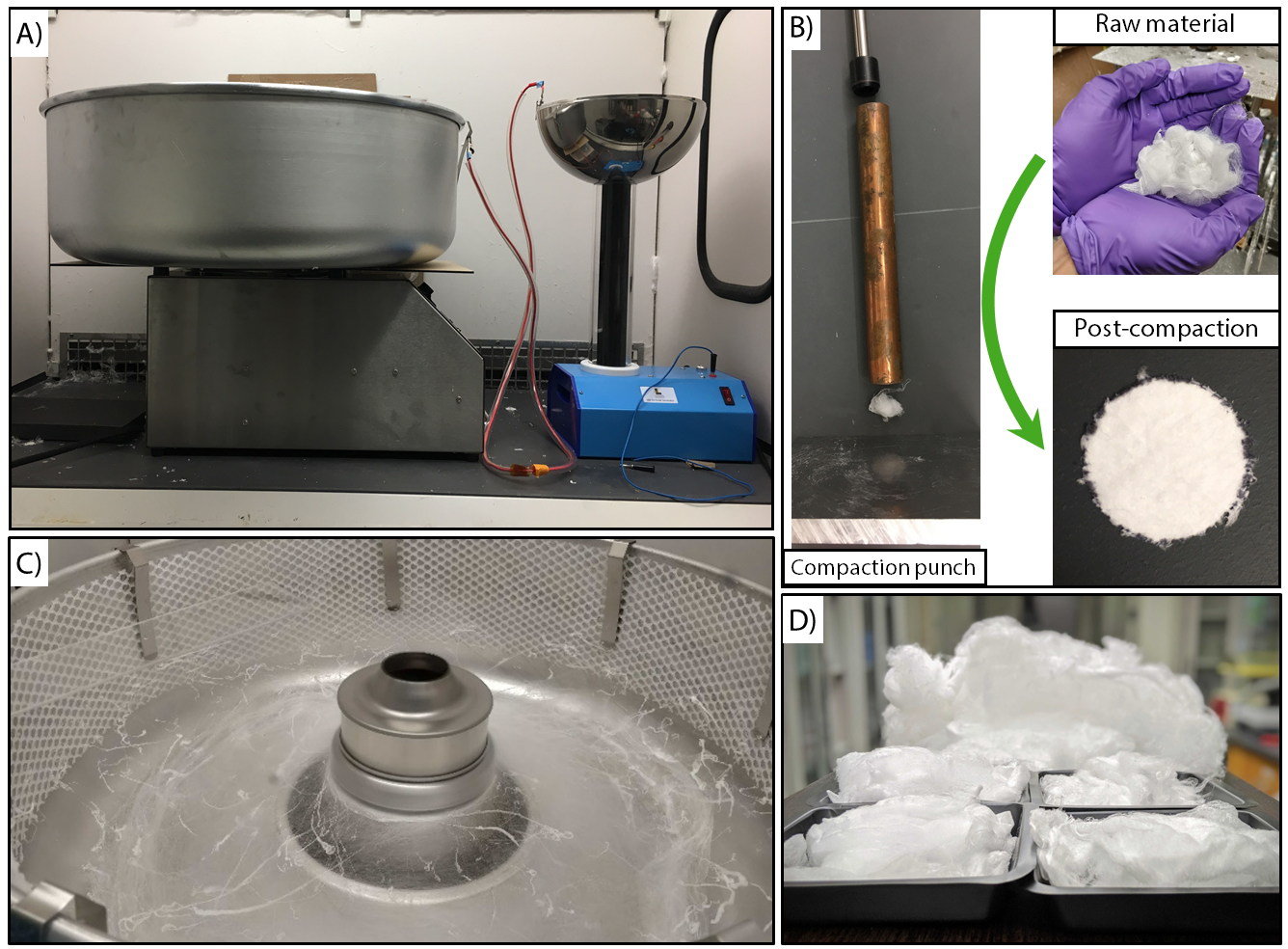}
\end{center}
\caption{  
	\textbf{Production of Filtration Media Using e-CMS.} \textbf{A} A commercial cotton candy machine is connected to a van de Graaff generator to produce a high voltage on the collection drum. \textbf{B} The raw material is collected as a loosely packed bale \textbf{right, top}. The raw material is subjected to compression \textbf{left} to produce samples of controlled grammage. Our highest performing samples gave a filtration efficiency equivalent to N95 \textbf{right, bottom}.  \textbf{C} the collection tub surrounds a spinneret that is filled with polypropylene resin and heated so that the resin flows through orifices along the perimeter of the spinneret. \textbf{D} The high-throughput nature of this process means bulk quantities of this material can be rapidly produced.
}
\label{fig:samplematerial}
\end{figure}

\subsection*{Process Design}
Centrifugal melt electrospinning of non-woven polypropylene fibers requires high temperatures (165\si{\degreeCelsius}), high rotational speeds (3,000 - 12,000 RPM) for producing fine fibers, and a means to impart electrostatic charge on the nascent fibers.
Commercial cotton candy machines, normally used for centrifugal spinning of sugar-based fibers, are a convenient apparatus for producing sufficiently high temperatures, although many designs do not have sufficiently fast rotational speeds. Larger industrial-grade machines, such as the one used in this report, achieve higher rotational speeds, enabling production of fibers suitable for use as a filtration material (\textbf{Figure \ref{fig:samplematerial}}). The material studied in this report was produced using a cotton candy machine modified only by replacing the standard aluminum mesh with a solid aluminum ring. The ring has several small apertures (600~\si{\micro\meter}) along its perimeter, thereby allowing more effective distribution of heat and better control over fiber morphology.

The choice of polymer resin is an important consideration for producing fibers. For example, previous work has shown a strong dependence of fiber diameter on melt flow rate, a measure closely related to molecular weight and viscosity \cite{raghavan2013}. In this study, we used three different types of polypropylene resin:  (1) high molecular weight isotactic (high-MW), (2) low molecular weight isotactic (low-MW), and (3) amorphous. We observed fiber formation for all three resins. The resulting fibers are collected as a loose bale, similar to how cotton candy appears. It is necessary to increase the density of the material before its filtration efficiency can be evaluated.

\begin{figure}[!ht]
\begin{center}
\includegraphics[width=0.75\linewidth]{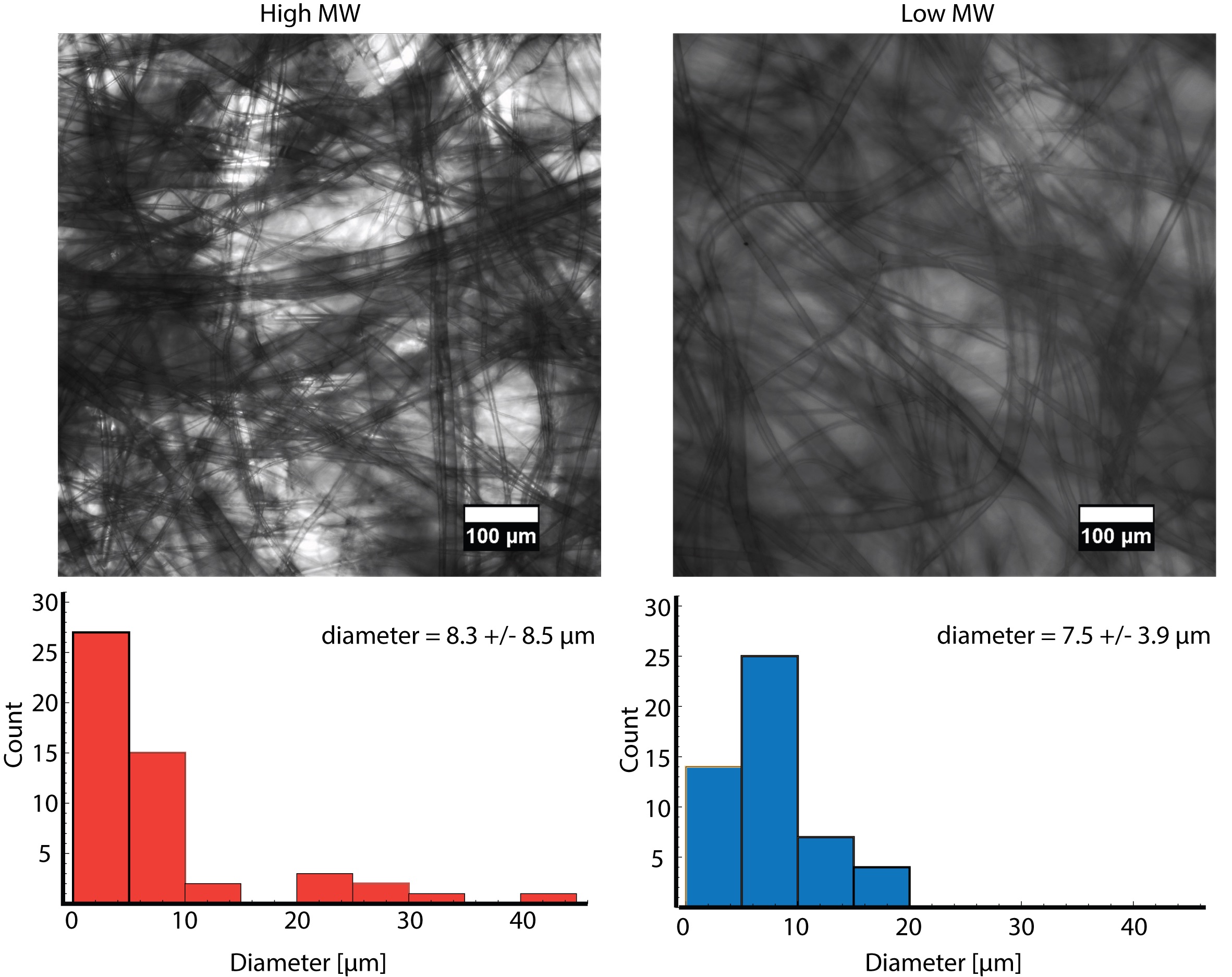}
\end{center}
\caption{  
	\textbf{Optical microscopy analysis.} (Top) confocal images of high (left) and low (right) MW fibers. Histogram (bottom) showing fiber diameters for the samples above. Average fiber diameter was calculated from N=50 unique fibers.
}
\label{fig:Lightmicro}
\end{figure}

\begin{figure}[!ht]
\begin{center}
\includegraphics[width=0.6\linewidth]{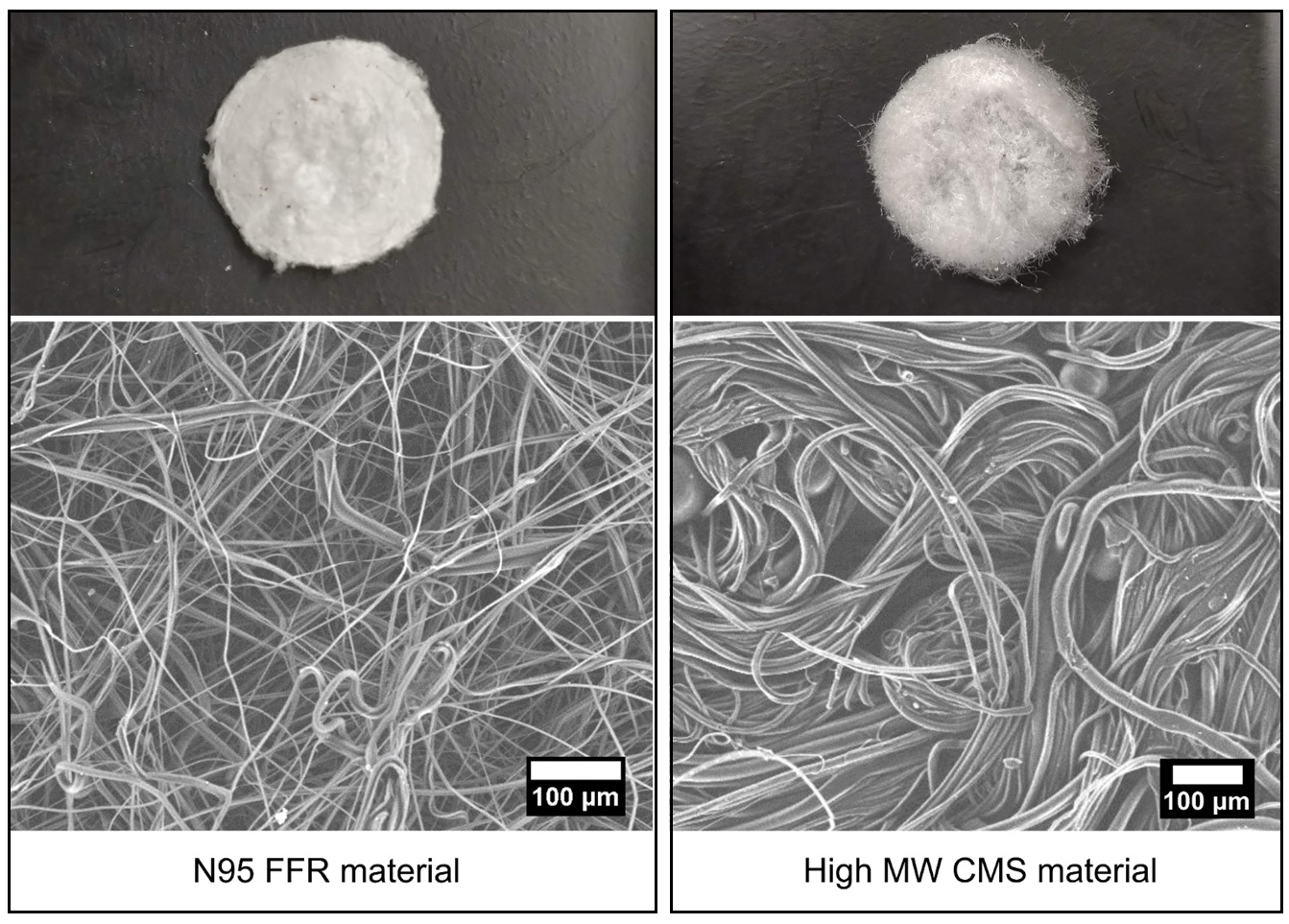}
\end{center}
\caption{  
	\textbf{Electron microscopy analysis.} SEM image taken for a N95 FFR (filtering facepiece respirator) material on the left; as compared to a high molecular weight CMS material on the right.
}
\label{fig:SEM}
\end{figure}

\subsection*{Material Characterization}
We hypothesized that the polymer properties of the polypropylene feedstock would significantly determine the material properties of filtration media produced by our CMS process. We found that material produced using isotactic polypropylene of low molecular weight (average $M_w$ $\sim$12,000, average $M_n$ $\sim$5,000) yielded fibers that were mechanically brittle and more prone to disintegration as compared to higher molecular weight isotactic polypropylene. Amorphous polypropylene produced dense material that displayed high cohesiveness and adhesiveness, meaning it was not compatible for application as FFR material. Confocal imaging of fibers produced from isotactic polypropylene revealed an average fiber diameter of around 7-8~\si{\micro\meter} (\textbf{Figure \ref{fig:Lightmicro}}), with a greater spread in fiber diameter in the case of high molecular weight polypropylene.  

After compaction, isotactic polypropylene samples exhibited grammages of about 620-695~\si{\gram\per\meter\squared}, significantly greater than commercial N95 FFR material (141~\si{\gram\per\meter\squared}). Compacted samples produced by CMS were also significantly thicker than commercial N95 FFR material, yielding samples 1.2-1.8~\si{\milli\meter} in thickness as compared to about 0.7~\si{\milli\meter}. Material density after compaction was also about twice that of commercial N95 FFR material (0.44 vs. 0.20~\si{\gram\per\meter\cubed}) (\textbf{Table \ref{table:1}}). Scanning electron microscopy (SEM) was used to compare a commercially available N95 filtration material with CMS-produced and compacted high molecular weight material (\textbf{Figure \ref{fig:SEM}}). The images indicate that the compaction results in the formation of a well-packed fiber matrix, suitable for further testing.

\begin{figure}[!ht]
\begin{center}
\includegraphics[width=0.7\linewidth]{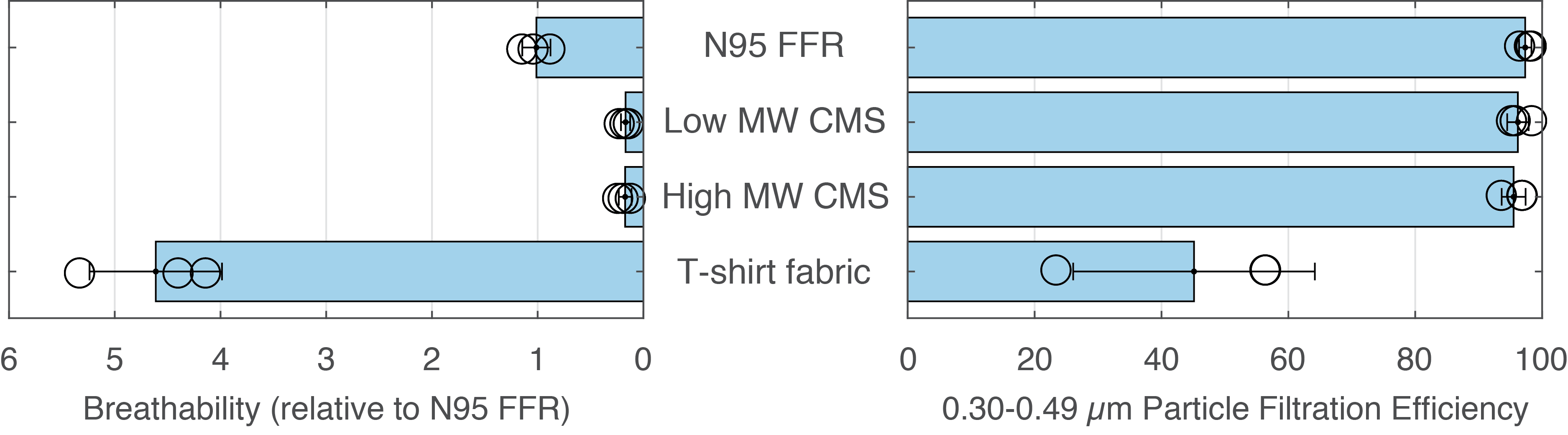}
\end{center}
\caption{  
	\textbf{Filtration efficiency and pressure drop testing of filtration material produced using CMS.} Particle filtration efficiency of filtration materials were tested using incense smoke as a source of particles and a handheld particle counter (Lighthouse 3016). Circular samples of commercial N95 FFR (Kimberly-Clark), compacted CMS material produced from polypropylene feedstock of two different molecular weights, and T-shirt fabric were excised and characterized in a custom-built filtration testing apparatus. Filtration efficiency was calculated as the ratio of detected particles (0.30-0.49~\si{\micro\meter}) with and without filter. T-shirt fabric material consisted of 50\% cotton, 25\% polyester, and 25\% rayon. Pressure drop testing was performed at a constant flow rate of 4.00~\si{\liter\per\minute} using an in-line pressure sensor (Honeywell) and the same excised material samples. Breathability is reported as the inverse of pressure drop across material and is taken relative to that measured for commercial N95 FFR material.
}
\label{fig:filtefficiency}
\end{figure}

\begin{figure}[!ht]
\begin{center}
\includegraphics[width=1.0\linewidth]{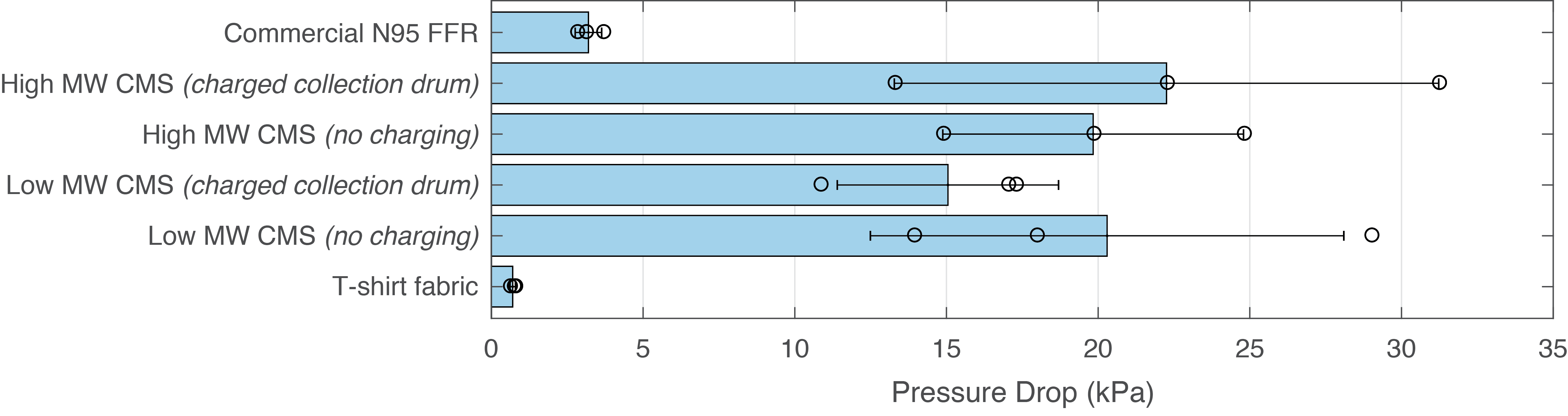}
\end{center}
\caption{
	\textbf{Pressure drop of all filtration samples tested.} 
}
\label{fig:PDall}
\end{figure}

Filtration media rated for N95-grade performance requires removal of at least 95\% of particles of average diameter 0.3~\si{\micro\meter}. To verify the applicability of our CMS material in the context of FFRs, we performed filtration testing using a custom-built setup involving a handheld particle counter and a capsule constructed from threaded PVC piping to hold circular samples excised from bulk, compacted CMS material. Using incense smoke as a source of particles primarily in the 0.3~\si{\micro\meter} diameter range, we found that material produced by our CMS process enabled filtration efficiencies that exceeded 95\% on average for three independently compacted samples. We found that N95-grade performance was achieved regardless of the molecular weight of polypropylene used (\textbf{Figure \ref{fig:filtefficiency}, right}). Samples excised from commercial N95 FFR material yielded an average filtration efficiency exceeding 97\%, as expected. By contrast, samples excised from T-shirt fabric material (suggested for homemade masks by the CDC) gave a filtration efficiency of 45\%, similar to previous reports \cite{mueller2018}. Airflow pressure drop across our CMS material samples was found to significantly greater than that of commercial N95 FFR samples (20 vs. 3.2~\si{\kilo\pascal}) (\textbf{Figures \ref{fig:filtefficiency}, left,} and \textbf{\ref{fig:PDall}}). This suggests that high filtration efficiency in the case of CMS material was achieved at the expense of breathability relative to commercially manufactured filtration material. The significantly higher grammage of compacted CMS samples relative to N95 FFR material likely gives rise to this reduced breathability and could be addressed by reducing the mass of material used for compaction. Moreover, given that increased electrostatic charging of non-woven fibrous material enables higher filtration efficiency for a given material density \cite{tsai2020}, introducing electrostatic charging could improve breathability of our CMS material while maintaining high filtration efficiency. 

\begin{table}[h!]
\centering
\begin{tabular}{||c | c c c ||} 
 \hline
Sample & Low MW CMS & High MW CMS & N95 FFR \\ [0.5ex] 
 \hline\hline
 A & 753.1 \textit{(0.440)} & 522.0 \textit{(0.446)} & 115.5 \textit{(0.165)} \\
 B & 629.0 \textit{(0.422)} & 564.8 \textit{(0.471)} & 145.5 \textit{(0.202)}\\
 C & 701.7 \textit{(0.465)} & 770.2 \textit{(0.412)} & 162.6 \textit{(0.246)}\\
 \hline  
 Grammage & 694.61$\pm$62.35 & 619.01$\pm$132.67 & 141.20$\pm$23.82  \\
 
 \textit{Density} & \textit{0.442}$\pm$\textit{0.021} &  \textit{0.443}$\pm$\textit{0.030} & \textit{0.204}$\pm$\textit{0.041}\\ 
 \hline
 
\end{tabular}
\caption{Grammages (in \si{\gram\per\meter\squared}) and densities (in ~\si{\gram\per\meter\cubed}) of compacted samples produced by centrifugal melt spinning of three different types of polypropylene feedstock}
\label{table:1}
\end{table}


\begin{table}[h!]
\centering
\begin{tabular}{||c | c c c | c||} 
 \hline
Material & Sample 1 & Sample 2 & Sample 3 & Average \\ [0.5ex] 
 \hline
 \hline
 Commercial N95 FFR & 97.76 & 98.04 & 96.22 & 97.34 \\
 \hline
 High MW CMS & 94.94 & 95.52 & 98.10 & 96.19 \\
 \hline
 Low MW CMS & 96.57 & 96.63 & 93.33 & 95.51 \\
 \hline  
 Woven T-shirt & 56.07 & 23.13 & 56.17 & 45.12  \\
 \hline
\end{tabular}
\caption{Filtration efficiency values of filtration material produced using CMS in comparison to other materials for particles in the range of 0.30-0.49~\si{\micro\meter}.}
\label{table:2}
\end{table}

\subsection*{Effects of Electrostatic Charging}

Although not necessary, electrostatic charging contributes significantly to filtration efficiency, especially for particles of diameter \si{<5\micro\meter}, a range potentially relevant to SARS-CoV-2 transmission \cite{Kowalski1999, van_doremalen_aerosol_2020}. We first attempted to introduce electrostatic charge on filtration material after production by CMS using corona charging\cite{Kao2004, tsai1993}, coupling to a van de Graaff generator, as well as triboelectric charge transfer using polystyrene material. A surface voltmeter was used to measure the voltage close to the insulator surfaces. Surface charge density on the fibers was then estimated based on the measured voltage readings (details in the methods section). Charge density on a commercial N95 FFR material was estimated to be around -2000~\si{\nano\coulomb\per\meter\squared}. Compacted filtration material produced using high MW polypropylene have exhibited charge densities close to -1000~\si{\nano\coulomb\per\meter\squared}. After charging using the above three methods, estimated surface charge densities on the CMS-produced fibers have been calculated to be as high as -9500~\si{\nano\coulomb\per\meter\squared}. However, our surface charge measurement method does not accurately reveal the homogeneity or stability of the transferred charges, and therefore careful interpretation of the measured charge density values is advised. Further measurements which can help analyze the uniformity of charges are necessary to inform how to use electrostatic charging for improved filtration efficiency.  

\begin{figure}[!ht]
\begin{center}
\includegraphics[width=0.6\linewidth]{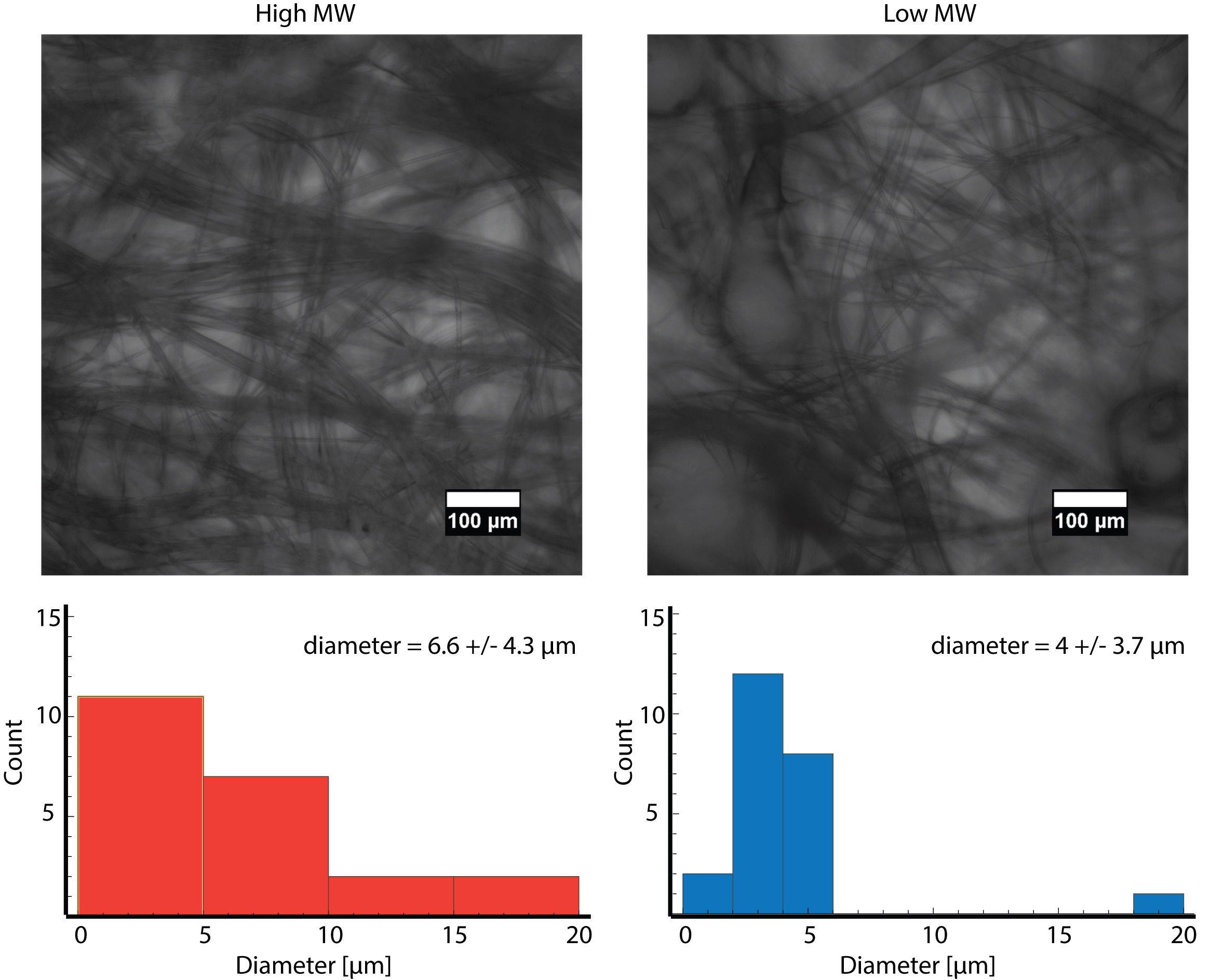}
\end{center}
\caption{  
	\textbf{Optical microscopy analysis.} (Top) confocal images of high (left) and low (right) MW fibers subject to a high voltage applied to the collection drum. Histogram (bottom) showing fiber diameters for the samples above. Average fiber diameter was calculated from N=50 unique fibers. 
}
\label{fig:charged}
\end{figure}

\begin{figure}[!ht]
\begin{center}
\includegraphics[width=1.0\linewidth]{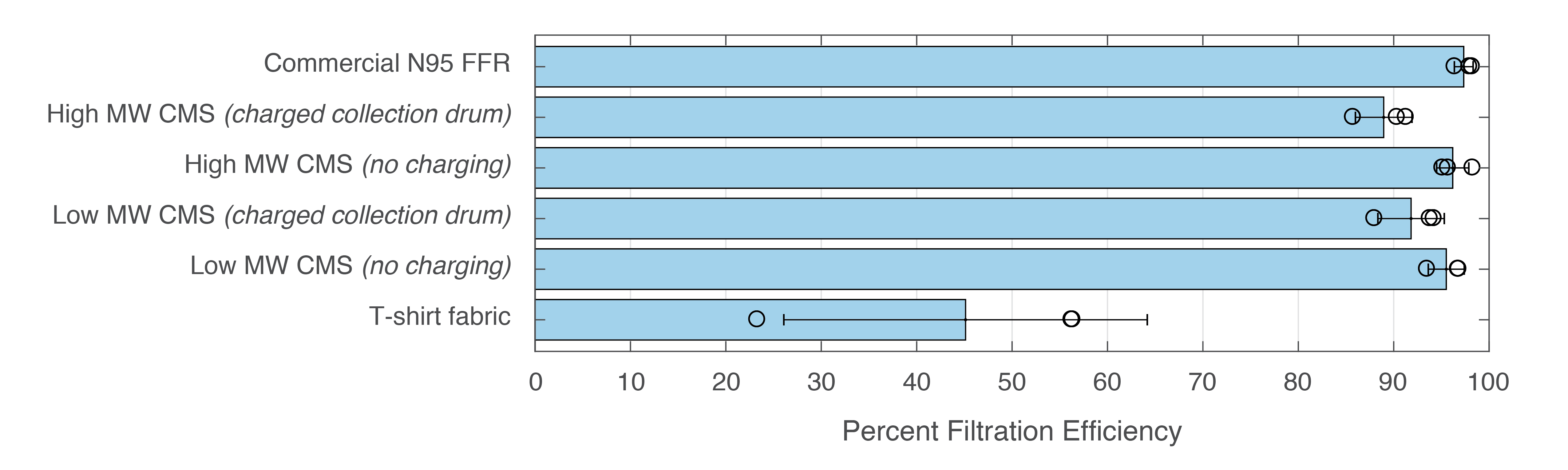}
\end{center}
\caption{  
	\textbf{Filtration efficiency of all filtration samples tested.} 
}
\label{fig:FEall}
\end{figure}

We also attempted charging of fibers during the production process (e-CMS) by applying an external electric field between the heated CCM spinneret (containing molten polypropylene) and collection drum. Charging of fibers would occur by trapping induced dipoles in the fibers in the aligned state after solidification \cite{Kao2004}. Preliminary results indicated that this method of charging did not noticeably influence fiber morphology or improve filtration efficiency (\textbf{Figures \ref{fig:charged} and \ref{fig:FEall}}). Parallel efforts by another group using a CMS process have similarly shown that kilovolt-strength electric potentials are not necessary (OIST) \cite{bandi2020}. Further development of the CMS process to accommodate and better understand potentially beneficial effects of charging during fiber production remains important.

\subsection*{Future Work and Process Scaling}
We are continuing to develop this approach to increase both output and quality at a lower cost. This involves not only continued tool building but also community engagement.

We are currently building a device from the group up using readily available components can reduce cost by avoiding the unnecessary, specialized components associated with a cotton candy machine. The motivation for this is twofold. First, a purpose built device will offer greater control over experimental parameters. For example, a simple spindle rotor offers greater control over and access to higher rotational speeds. This has the potential to reduce fiber diameter, offering improvements in filtration performance, and increasing throughput. Additionally, a modular device design will allow for much more flexibility in testing different approaches for charging the material during production. This is especially valuable since it allows for community based development, where improvements to the process can be made in a distributed way. Second, the use of readily available components can reduce cost by up to 50\%. The current unit cost of our prototype is $\sim\$1,000$. This cost is based on a standard cotton candy machine (\$650), a custom machined cylinder (\$50), and a van de Graaff Generator (\$300). A 50\% reduction in unit cost increases the economic scalability of this approach.

The two main objectives of this work is to develop an FFR manufacturing method in an affordable and also high-throughput manner. During our prototyping phase, we iterated over several unique CMS designs and have been able to consistently produce at least 1-2~\si[per-mode=symbol]{\gram\per\minute} of material across a range of processing parameters. Given that a typical N95 FFR contains \si{\sim 2 \gram} of filtration material, we can gain some perspective on the extent to which the present proposal can make a meaningful impact. A single CMS  can produce enough filtration material for 1000 FFRs in a day. Therefore, a small scale facility consisting of 6 CMS devices operated for 12 hours by a small group of people to produce enough material for 5000 FFRs per day. This is sufficient to supply the daily demand of a large medical facility \cite{ppe2016}. Alternatively, 1000 CMS devices operating as a distributed network provides a rapidly configurable and resilient manufacturing capacity equivalent to a single, industrial-scale facility. We believe that the approach described here has the potential to make a significant contribution towards addressing the current FFR shortage. Our work represents a starting point for others to construct and develop their own affordable modular processes for producing high quality filtration material.

\section*{Methods}

\subsection*{Materials}
Amorphous polypropylene, isotactic low molecular weight (average $M_w$ $\sim$12,000, average $M_n$ $\sim$5,000) polypropylene, and isotactic high molecular weight (average $M_w$ $\sim$250,000, average $M_n$ $\sim$67,000) were purchased from Sigma-Aldrich (St. Louis, MO). Commercial N95 FFR material was obtained from a Kimberly-Clark 62126 Particulate Filter Respirator and Surgical Mask (Kimberly-Clark Professional, Roswell, GA).

\subsection*{Preparation of filtration material}
Nano- and microfibers were prepared using a modified cotton candy machine (Spin Magic 5, Paragon, USA). Polypropylene resin was placed directly into the preheated spinneret while in motion. Fibrous material was collected on the machine collection drum and compacted against a heated metal plate (130\si{\degreeCelsius}, 30~\si{\second}, 4.23~\si{\kg\per\cm\squared}) using a cylindrical pipe and plunger. Three independent replicates were compacted for each sample tested, sourcing from the same batch of material produced for each of the three types of polypropylene. Circular samples (17.25~\si{\mm} diam.) for filtration testing were excised from this compacted material. 

\begin{figure}[!ht]
\begin{center}
\includegraphics[width=\linewidth]{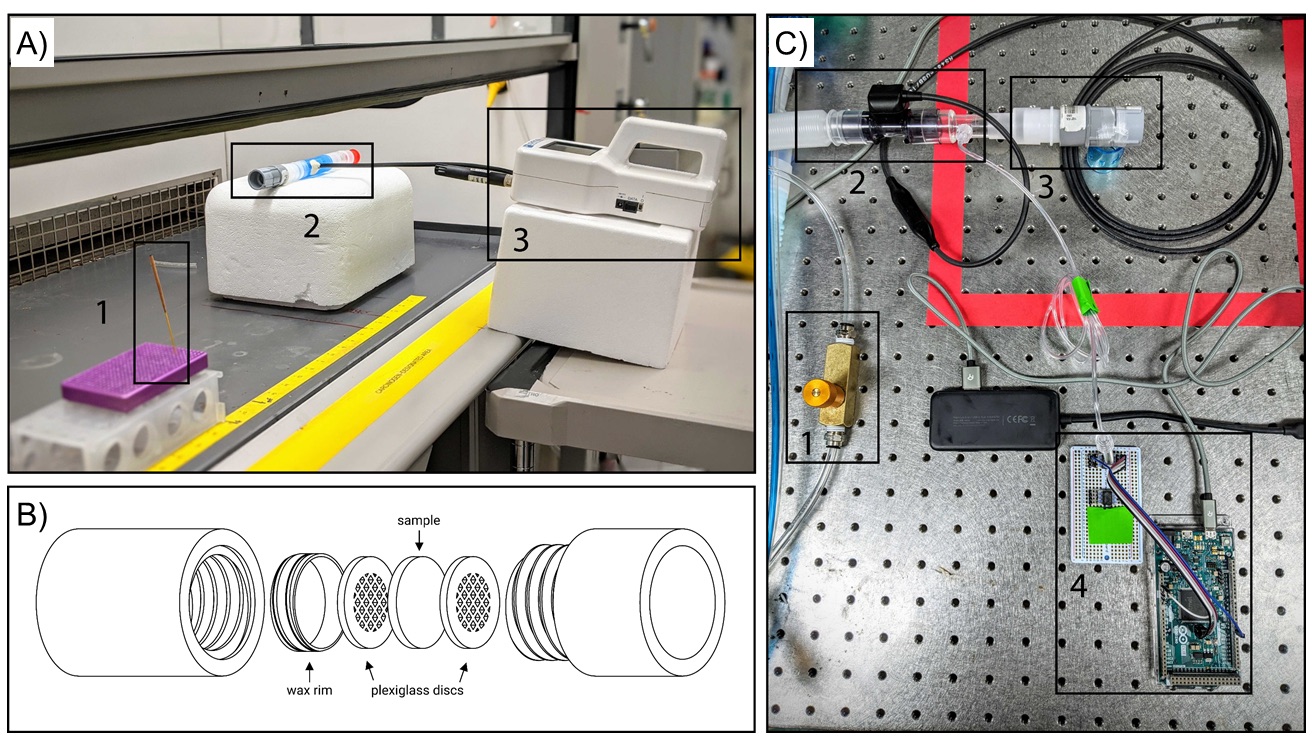}
\end{center}
\caption{  
	\textbf{A. Filtration testing setup} 1: Incense stick 2: Test filter assembly. 3: Lighthouse 3016 handheld particle counter
	\textbf{B. Test filter assembly} Compressed sample is placed between two acrylic mesh screens, sealed on the sides with paraffin tape and held in place using the pipe screw setup.
	\textbf{C. Pressure drop testing} 1: Flow control valve 2: Airflow measurement sensor 3: Test filter assembly 4: Pressure sensor and micro-controller.  
}
\label{fig:testing}
\end{figure}

As shown in Figure~\ref{fig:testing} (B), the circular samples are placed into a test filter assembly.
A test filter consists of a disc of sample material (1~\si{\mm} thick) held between two laser-cut Plexiglas mesh screens (1.6~\si{\mm} thick, 17.25~\si{\mm} diam.) (see Figure~\ref{fig:testing}).
The filter assembly is inserted between two threaded PVC pipe connectors that are press-fit onto the testing jig. The  edges of the filter assembly are wrapped with a paraffin wax seal (Parafilm, Bemis Inc, USA) to secure the three layers together and prevent air from leaking around the filter within the PVC pipe.

\subsection*{Electron microscopy}
A Hitachi TM-1000 tabletop SEM was used to obtain the micrographs. The fiber samples were attached to the stage using conductive silver paste. No sputter coat was added to the materials.

\subsection*{Electrostatic charging}
Corona charging was done by placing samples on top of a grounded aluminum plate and applying a steady ion current through point electrodes that are connected to a high DC voltage source (Model PMT2000, Advanced Research Instruments Corp.). The distance between the point electrodes and the sample was about 5~\si{\mm}. The voltage of the emitters was set to 4800~\si{\volt}. Charging beyond 30 minutes brought no additional increase in surface charge. E-CMS was performed by connecting the collection drum with a Van De Graaff generator, resulting in a steady-state voltage of approximately -18~\si{\kV}. Triboelectric charging was performed by rubbing the sample directly against a range of materials. In particular, metal and cardboard were found capable of imparting negative charges on polypropylene samples, while polystyrene sheets had the opposite effect.

\subsection*{Charge measurements}
A surface DC voltmeter (SVM2, Alphalab Inc., USA) was used to measure the voltage of the fiber material, 2.54~\si{\cm} away from the fiber mesh surface. The surface charge density is estimated using manufacturer-provided\cite{svm2} equations
\[
    \begin{array}{ll} 
    \frac{Q}{A} &= \alpha V f(f-1) \\
    f &= \sqrt{1+\frac{D^2}{4L^2}} ,
    \end{array}
\]
where
$Q$ is the surface charge (\si{\coulomb}), 
$A$ is the area (\si{\cm\squared}) of fiber mesh, 
$V$ is the measured voltage (\si{\volt}), 
$D$ is the diameter (\si{\cm}) of the fiber mesh, 
$L$ is the distance (\si{\cm}) of the mesh surface away from voltmeter sensor,
and $\alpha = 3.6\times 10^{-14} $ is device-specific parameter provided by the manufacturer.

\subsection*{Filtration testing}
The filter efficiency testing is done using a custom experimental setup that includes a handheld particle counter (Model 3016 IAQ, LightHouse, USA), 100~\si{\gram} incense (Nag Champa, Satya Sai Baba, India), and connectors (universal cuff adaptor, teleflex multi-adaptor). Whereas a typical testing setup uses an all-in-one filter tester, \emph{e.g.} an 8130A automated filter tester (TSI Automated, USA) that supports a flow rate up to 110~\si{\liter\per\minute}, our system was run at an airflow rate of 2.83~\si{\liter\per\minute}. The incense produces particles of various sizes, including those in the range picked up by the detector (0.30-10~\si{\micro\meter}), and primarily in the 0.30-0.49~\si{\micro\meter} range. To calculate the filtration efficiency, the ratio of unfiltered particles detected to the number of particles detected without filter is subtracted from unity. 

\subsection*{Pressure drop testing}

The same sample-containing capsule used for filtration testing was also used for pressure drop measurements. Compressed air flow was delivered at a constant rate of 4.00~\si{\liter\per\minute}, similar to that experienced during human respiration accounting for the smaller sample cross-sectional area compared to a full FFR. The airflow rate is measured using a Mass Flow Meter SFM3300 (Sensirion AG, Switzerland) and the pressure drop is measured using a Honeywell Trustability Series pressure sensor (Model HSCDANN005PGSA3, Honeywell International Inc., USA). Sensor data is acquired using an Arduino Mega microcontroller development board (Arduino AG, Italy).

\bibliography{references}

\begin{thebibliography}{10}
\urlstyle{rm}
\expandafter\ifx\csname url\endcsname\relax
  \def\url#1{\texttt{#1}}\fi
\expandafter\ifx\csname urlprefix\endcsname\relax\def\urlprefix{URL }\fi
\expandafter\ifx\csname doiprefix\endcsname\relax\def\doiprefix{DOI: }\fi
\providecommand{\bibinfo}[2]{#2}
\providecommand{\eprint}[2][]{\url{#2}}

\bibitem{national1996niosh}
\bibinfo{author}{{National Institute for Occupational Safety and Health}}.
\newblock \bibinfo{title}{{NIOSH guide to the selection and use of particulate
  respirators certified under 42 CFR 84}} (\bibinfo{year}{1996}).
\newblock
  \bibinfo{note}{\url{https://www.cdc.gov/niosh/docs/96-101/default.html}}.

\bibitem{kim2007direct}
\bibinfo{author}{Kim, J.}, \bibinfo{author}{Jasper, W.} \&
  \bibinfo{author}{Hinestroza, J.}
\newblock \bibinfo{journal}{\bibinfo{title}{Direct probing of solvent-induced
  charge degradation in polypropylene electret fibres via electrostatic force
  microscopy}}.
\newblock {\emph{\JournalTitle{Journal of Microscopy}}}
  \textbf{\bibinfo{volume}{225}}, \bibinfo{pages}{72--79}
  (\bibinfo{year}{2007}).

\bibitem{kim2010application}
\bibinfo{author}{Kim, J.}, \bibinfo{author}{Jasper, W.},
  \bibinfo{author}{Barker, R.} \& \bibinfo{author}{Hinestroza, J.}
\newblock \bibinfo{journal}{\bibinfo{title}{Application of electrostatic force
  microscopy on characterizing an electrically charged fiber}}.
\newblock {\emph{\JournalTitle{Fibers and Polymers}}}
  \textbf{\bibinfo{volume}{11}}, \bibinfo{pages}{775--781}
  (\bibinfo{year}{2010}).

\bibitem{bonilla2012direct}
\bibinfo{author}{Bonilla, R.}, \bibinfo{author}{Avila, A.},
  \bibinfo{author}{Montenegro, C.} \& \bibinfo{author}{Hinestroza, J.}
\newblock \bibinfo{journal}{\bibinfo{title}{Direct observation of the spatial
  distribution of charges on a polypropylene fiber via electrostatic force
  microscopy}}.
\newblock {\emph{\JournalTitle{Journal of microscopy}}}
  \textbf{\bibinfo{volume}{248}}, \bibinfo{pages}{266--270}
  (\bibinfo{year}{2012}).

\bibitem{huang2017}
\bibinfo{author}{Huang, T.}, \bibinfo{author}{Lim, H.~S.} \&
  \bibinfo{author}{Yung, W.-S.}
\newblock \bibinfo{title}{Electret nanofibrous web as air filtration media}
  (\bibinfo{year}{U.S. Patent 9 610 588 B2, 2019}).

\bibitem{huang2019}
\bibinfo{author}{Huang, T.}, \bibinfo{author}{Croft, J.} \&
  \bibinfo{author}{Dilworth, Z.~R.}
\newblock \bibinfo{title}{Melt spin filtration media for respiratory devies and
  face masks} (\bibinfo{year}{U.S. Patent 10 456 724 B2, 2019}).

\bibitem{drabek2019}
\bibinfo{author}{Drabek, J.} \& \bibinfo{author}{Zatloukal, M.}
\newblock \bibinfo{journal}{\bibinfo{title}{Meltblown technology for production
  of polymeric microfibers/nanofibers: A review.}}
\newblock {\emph{\JournalTitle{Physics of Fluids}}}
  \textbf{\bibinfo{volume}{31}},
  \doiprefix\url{https://doi.org/10.1063/1.5116336} (\bibinfo{year}{2019}).

\bibitem{tsai2020}
\bibinfo{author}{Tsai, P.~P.}
\newblock \bibinfo{howpublished}{Personal communication}.

\bibitem{klaase1984}
\bibinfo{author}{Klaase, P. T.~A.} \& \bibinfo{author}{van Turnhout, J.}
\newblock \bibinfo{title}{Method for manufacturing an electret filter medium}
  (\bibinfo{year}{U.S. Patent 4 588 537 A, 1984}).

\bibitem{reifenhauser2020}
\bibinfo{author}{Reifenh{\"a}user}.
\newblock \bibinfo{title}{{Reifenhäuser Reicofil Shortens Delivery Time for
  the Supply of Meltblown Lines}} (\bibinfo{year}{2020}).
\newblock
  \bibinfo{note}{\url{https://www.reifenhauser.com/en/news/reifenhaeuser_reicofil_shortens_delivery_time_for_the_supply_of_meltblown_lines}}.

\bibitem{rogalski2017}
\bibinfo{author}{Rogalski, J.~J.}, \bibinfo{author}{Bastiaansen, C. W.~M.} \&
  \bibinfo{author}{Peijs, T.}
\newblock \bibinfo{journal}{\bibinfo{title}{Rotary jet spinning review – a
  potential high yield future for polymer nanofibers}}.
\newblock {\emph{\JournalTitle{Nanocomposites}}} \textbf{\bibinfo{volume}{3}},
  \bibinfo{pages}{97–121}, \doiprefix\url{10.1080/20550324.2017.1393919}
  (\bibinfo{year}{2017}).

\bibitem{raghavan2013}
\bibinfo{author}{Raghavan, B.}, \bibinfo{author}{Soto, H.} \&
  \bibinfo{author}{Lozano, K.}
\newblock \bibinfo{journal}{\bibinfo{title}{Fabrication of melt spun
  polypropylene nanofibers by forcespinning}}.
\newblock {\emph{\JournalTitle{Journal of Engineered Fibers and Fabrics}}}
  \textbf{\bibinfo{volume}{8}},
  \doiprefix\url{https://doi.org/10.1177/155892501300800106}
  (\bibinfo{year}{2013}).

\bibitem{parker2010}
\bibinfo{author}{Badrossamay, M.~R.}, \bibinfo{author}{McIlwee, H.~A.},
  \bibinfo{author}{Goss, J.~A.} \& \bibinfo{author}{Parker, K.~K.}
\newblock \bibinfo{journal}{\bibinfo{title}{Nanofiber assembly by rotary
  jet-spinning}}.
\newblock {\emph{\JournalTitle{Nano Letters}}} \textbf{\bibinfo{volume}{10}},
  \bibinfo{pages}{2257–2261},
  \doiprefix\url{https://doi.org/10.1021/nl101355x} (\bibinfo{year}{2010}).

\bibitem{mueller2018}
\bibinfo{author}{Mueller, W. e.~a.}
\newblock \bibinfo{journal}{\bibinfo{title}{The effectiveness of respiratory
  protection worn by communities to protect from volcanic ash inhalation. part
  i: Filtration efficiency tests}}.
\newblock {\emph{\JournalTitle{International Journal of Hygiene and
  Environmental Health}}} \textbf{\bibinfo{volume}{6}},
  \bibinfo{pages}{97–121},
  \doiprefix\url{https://doi.org/10.1016/j.ijheh.2018.03.0129}
  (\bibinfo{year}{2018}).

\bibitem{Kowalski1999}
\bibinfo{author}{Kowalski, W.}, \bibinfo{author}{Bahnfleth, W.} \&
  \bibinfo{author}{Whittam, T.}
\newblock \bibinfo{journal}{\bibinfo{title}{Filtration of airborne
  microorganisms: Modeling and prediction}}.
\newblock {\emph{\JournalTitle{ASHRAE Trans.}}} \textbf{\bibinfo{volume}{105}}
  (\bibinfo{year}{1999}).

\bibitem{van_doremalen_aerosol_2020}
\bibinfo{author}{van Doremalen, N.} \emph{et~al.}
\newblock \bibinfo{journal}{\bibinfo{title}{Aerosol and {Surface} {Stability}
  of {SARS}-{CoV}-2 as {Compared} with {SARS}-{CoV}-1}}.
\newblock {\emph{\JournalTitle{New England Journal of Medicine}}}
  \textbf{\bibinfo{volume}{382}}, \bibinfo{pages}{1564--1567},
  \doiprefix\url{10.1056/NEJMc2004973} (\bibinfo{year}{2020}).

\bibitem{Kao2004}
\bibinfo{author}{Kao, K.~C.}
\newblock \bibinfo{title}{5 - electrets}.
\newblock In \bibinfo{editor}{Kao, K.~C.} (ed.)
  \emph{\bibinfo{booktitle}{Dielectric Phenomena in Solids}},
  \bibinfo{pages}{283 -- 326} (\bibinfo{publisher}{Academic Press},
  \bibinfo{address}{San Diego}, \bibinfo{year}{2004}).

\bibitem{tsai1993}
\bibinfo{author}{Tsai, P.~P.} \& \bibinfo{author}{Wadsworth, L.~C.}
\newblock \bibinfo{title}{Method and apparatus for the electrostatic charging
  of a web or film} (\bibinfo{year}{U.S. Patent 5 401 446 A, 1993}).

\bibitem{bandi2020}
\bibinfo{author}{Bandi, M.~M.}
\newblock \bibinfo{title}{N95-electrocharged filtration principle based face
  mask design using common materials} (\bibinfo{year}{2020}).
\newblock \bibinfo{note}{\url{https://groups.oist.jp/nnp/diy-face-mask}}.

\bibitem{ppe2016}
\bibinfo{author}{{Center for Disease Control}}.
\newblock \bibinfo{title}{Estimated personal protective equipment needed for
  healthcare facilities} (\bibinfo{year}{2016}).
\newblock
  \bibinfo{note}{\url{https://www.cdc.gov/vhf/ebola/healthcare-us/ppe/calculator.html}}.

\bibitem{svm2}
\bibinfo{author}{{Alphalab Inc.}}
\newblock \bibinfo{title}{{Surface DC voltmeter model SVM2 quick start
  instructions }} (\bibinfo{year}{2018}).
\newblock
  \bibinfo{note}{\url{https://www.alphalabinc.com/wp-content/uploads/2018/03/SVM2-2013.pdf}}.

\end{thebibliography}

\subsection*{Acknowledgements}
The authors are greatful to Dr. George Herring, Hongquan Li, Prof. Anna Paradowska of the University of Sydney, and Tyler Orr for helpful discussion relating to project design, parts machining and material characterization. The authors thank Hongquan Li for assistance in pressure drop experiments and Prof. Fabian Pease of the Electrical Engineering Department at Stanford University for assistance in SEM characterization. The authors also deeply thank Edward Mazenc, Yuri Lensky, Daniel Ranard, and Abby Kate Grosskopf for contributing in the process design process. The authors would like to thank financial support from UCSF COVID-19 Response Fund, Schmidt Futures, Moore Foundation, CZ BioHub, NSF CCC grant (DBI 1548297) and HHMI-Gates Faculty Award. 


\newpage
\section*{Supplementary Information}

\subsection*{Distributed manufacturing}

\begin{figure}[!ht]
\begin{center}
\includegraphics[width=0.6\linewidth]{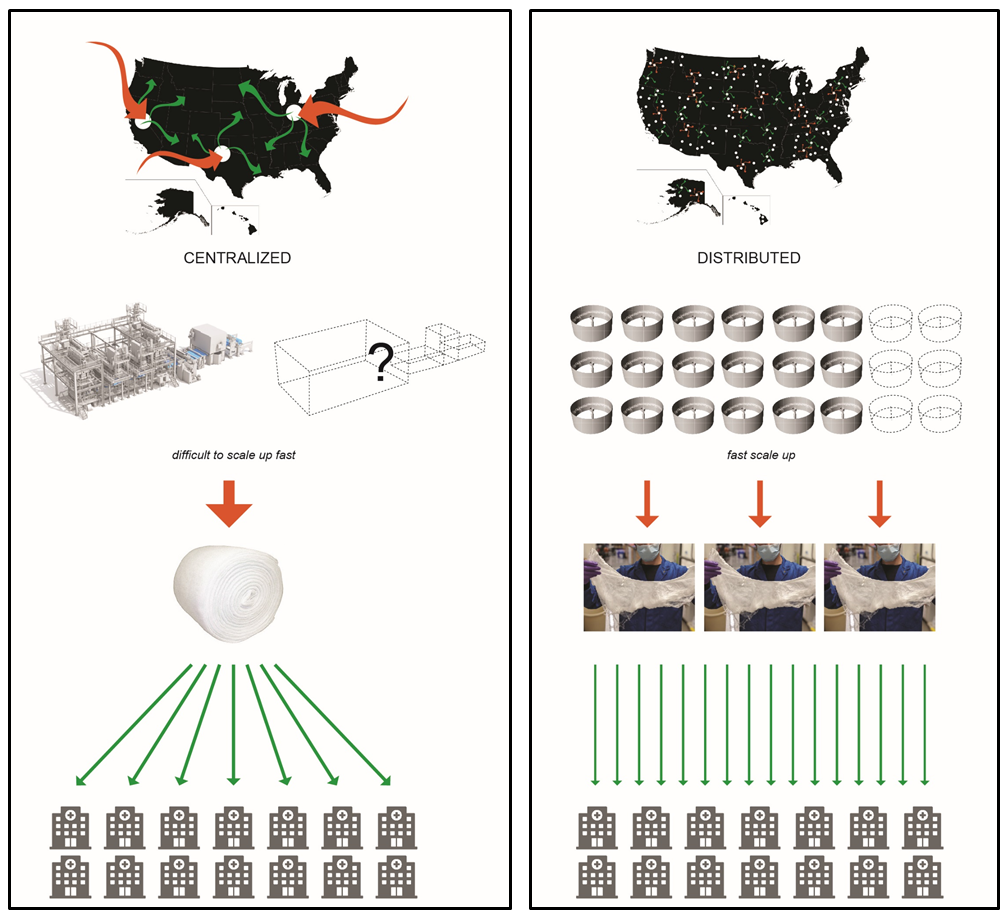}
\end{center}
\caption{  
	\textbf{Centralized vs. Distributed Manufacturing} 
}
\label{fig:distributed}
\end{figure}

\subsection*{Results of charging on fiber morphology}

\subsection*{Full filtration efficiency and pressure drop measurement results}

\begin{figure}[!ht]
\begin{center}
\includegraphics[width=0.75\linewidth]{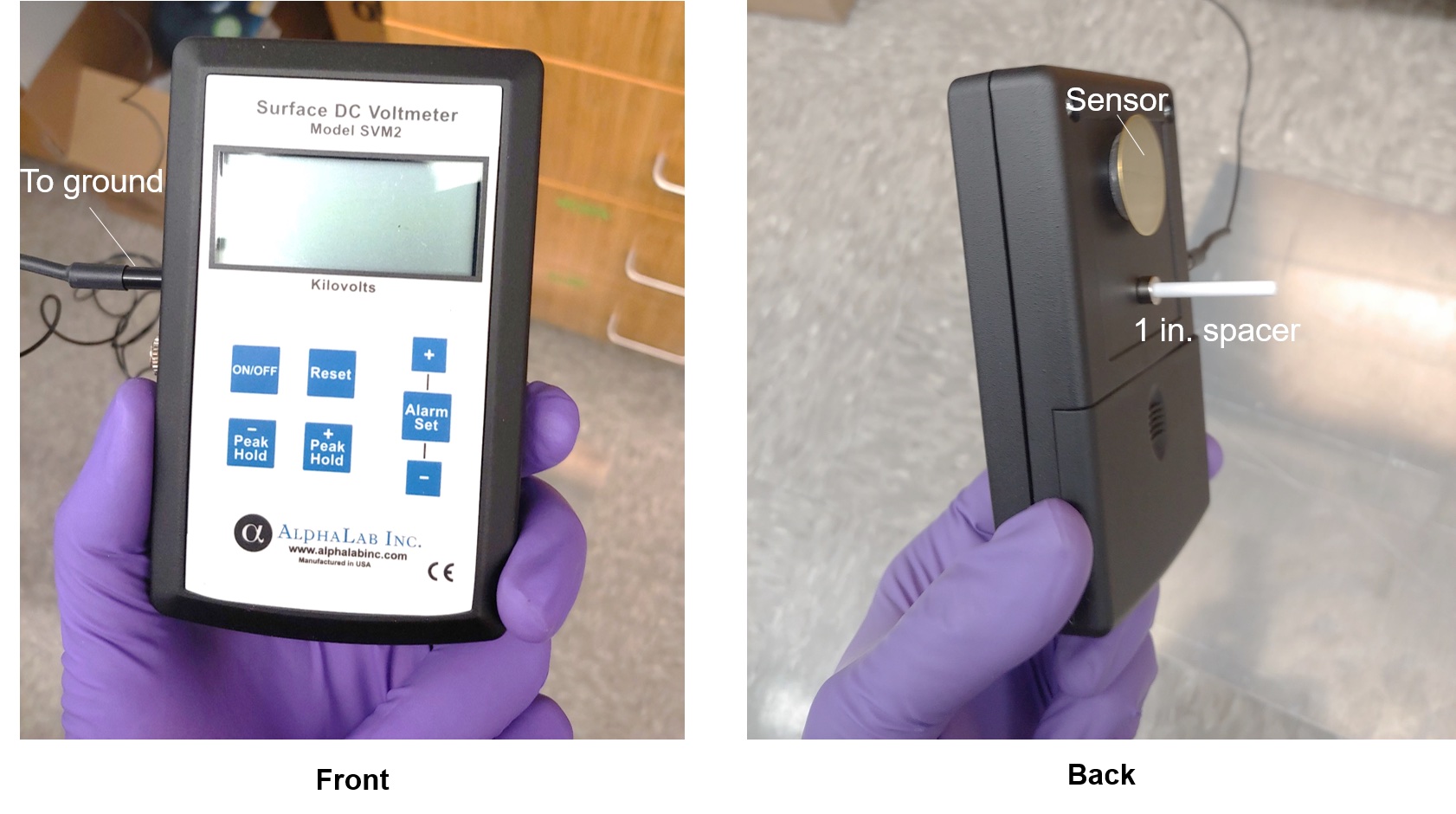}
\end{center}
\caption{
	\textbf{Surface charge measurement.} A commercial voltmeter (alphalabs inc.) was used to measure the surface voltage of sample materials. The surface charges were then calculated based on the measured voltages. Samples were kept 2.54~\si{\cm} away from the sensor for the measurements. 
}
\label{fig:chargemeter}
\end{figure}

\end{document}